\documentclass[prb,twocolumn]{revtex4}
\usepackage{epsf}
\begin{document}
\title{Gravity, Bose-Einstein Condensates and Gross-Pitaevskii Equation}
\author{Patrick Das Gupta}
\affiliation{Department of Physics and Astrophysics, University of Delhi, Delhi - 110 007 (India)}
\email{patrick@srb.org.in}

\begin{abstract}
We explore the effect of mutual gravitational interaction between ultra-cold gas atoms on the dynamics of Bose-Einstein condensates (BEC). Small amplitude oscillation of BEC is studied by applying variational technique to reduce the Gross-Pitaevskii equation, with gravity included, to the equation of motion of a  particle moving in a potential. According to our analysis, if the s-wave scattering length can be tuned to zero using Feshbach resonance for future BEC with occupation numbers as high as $\approx 10^{20}$,    there exists a critical ground state occupation number above which the BEC is  unstable, provided that its constituents interact with a $1/r^3 $ gravity at short scales.
\end{abstract}


\maketitle

\section{INTRODUCTION}
Gravity is the weakest of all forces. This is essentially due to the smallness of Newton's gravitational constant 
(or, equivalently, largeness of Planck mass),   measured on scales larger than tens of km \cite{bec0}. However, to resolve issues pertaining to naturalness and  hierarchy  problems in the Standard Model of particle physics, it has been conjectured that if  large extra dimensions exist, the effective gravitational coupling strength  can be larger  at sub-mm scales\cite{bec1,bec2}.  With the advent of  exciting precision experiments involving Bose-Einstein condensation of alkali atoms and molecules at ultra-low temperatures \cite{bec3,bec4}, it is  but natural to study  effects of enhanced gravity ensuing from large extra dimensions (LED) on such macroscopic quantum phenomena. 

In this context, Dimopoulos and Geraci have proposed an interesting experiment to probe gravity at sub-micron scale through measurements of relative phase evolution rates in Bose-Einstein condensates (BEC)  prepared in coherent superposition of states localized at two distinct potential wells, both situated near a moving wall of alternating gold and silver metal objects that form a periodic  massive source of gravity  \cite{bec5}. Similarly, Sigurdsson has suggested measuring fringe shifts of interfering pair of BEC  
 falling past a long and narrow cylindrical mass in order to estimate  modified transverse gravitational acceleration, provided that the LED sub-mm scale is in excess of 0.01 mm \cite{bec12}.
 
Interestingly  enough, the typical separation between  atoms in ultra-cold gases is only about few 100 nm.  This induces one to explore  effects of mutual gravitational interaction between  individual atoms of a BEC  on its quantum dynamics, and ask whether such weak but long range forces can lead to instabilities. In this paper, we carefully examine some  aspects of these ideas using variational method.
 
\section{Gross-Pitaevskii Equation and Large Extra Dimensions induced Gravity}

For $N$ identical bosons constituting a dilute BEC at temperature  $ T \approx 0^\circ  K $, the many body wavefunction $ \Psi (\vec{r_1}, \vec{r_2},..,\vec{r_N})$  describing the condensate can be expressed upto a good approximation (assuming that the bosons interact weakly with each other) as,
$$ \Psi (\vec{r_1}, \vec{r_2},..,\vec{r_N}) \cong  \prod ^N _{j=1} \psi (\vec{r_j}) $$
where $\psi (\vec{r})$ is the normalized ground state wavefunction for a single boson. As  each boson, in this case, is approximately in the same state,  $\psi (\vec{r})$ acts as the condensate wavefunction.

In the  $ T = 0^\circ  K $ mean field approximation, dynamical evolution of the condensate wavefunction  $\psi (\vec{r},t)$ (normalized to unity)  is, to a good extent, governed by the Gross-Pitaevskii equation,
$$ i \hbar {{\partial \psi}\over{\partial t}}  = \bigg [- \frac {\hbar^2}{2m} \nabla ^2    + V_{ext} + N \int {V(\vec{r} - \vec{u}) \vert \psi (\vec{u},t)\vert ^2 d^3u} \bigg ] \psi (\vec{r},t) \eqno(1)$$
where $m$, $V_{ext}(\vec{r})$ and $V (\vec{r})$ are the boson mass, the trap potential energy required to   confine the BEC and the interaction potential energy between two bosons, respectively. 

For the present purpose, the interaction potential energy $V$ in eq.(1) is a combination of s-wave scattering potential and the
 inter-bosonic gravitational potential energy $V_g$, so that,
 $$ V(\vec{r} - \vec{u})= \frac {4 \pi \hbar ^2  a} {m} \delta ^3 (\vec{r} - \vec{u}) + V_g (\vert \vec{r} - \vec{u} \vert)\ .\eqno(2) $$
 where $a$ is the s-wave scattering length.
 
 Substitution of eq.(2) in eq.(1) results in the standard Gross-Pitaevskii equation (GPE) \cite{bec4},
$$i \hbar {{\partial \psi}\over{\partial t}} = \bigg [- \frac {\hbar^2}{2m} \nabla ^2    + V_{ext} + N g  \vert \psi (\vec{r},t)\vert ^2 + $$
$$  + N \int {V_g (\vert \vec{r} - \vec{u} \vert) \vert \psi (\vec{u},t)\vert ^2 d^3u} \bigg ] \psi (\vec{r},t) \eqno(3)$$
where  $g \equiv \frac {4 \pi \hbar ^2  a} {m}$. 

It is interesting to note  that the quantum dynamics of a BEC, comprised of ultra-cold bosonic atoms anchored to a planar honeycomb optical lattice and  interacting weakly to one another via  a contact interaction  much like the first term of the RHS of eq.(2), is   described by a nonlinear Dirac equation \cite{bec6}. Furthermore, the pseudospin degrees of freedom associated in this case with the two inequivalent  sites of the sublattice display  half integral spin angular momentum features, stretching the graphene analogy farther, even though the system  is a bosonic one \cite{bec7}.
 
 
 The GPE of eq.(3) can easily be derived from the following action  by demanding it to be  stationary under infinitesimal variations of $\psi $ and $\psi^*$,
 $$ S = \int {dt \int {d^3r \ \mathcal{L}}} \eqno(4)$$
 where the Lagrangian density $ \mathcal{L} $ is given by,
 $$ \mathcal{L}= \frac {i \hbar}{2} \bigg \lbrace  \psi {{\partial \psi^*}\over{\partial t}} - \psi^* {{\partial \psi}\over{\partial t}}\bigg \rbrace + \frac {\hbar^2}{2m}   \nabla \psi^*. \nabla \psi + V_{ext}\ \vert \psi \vert ^2 + $$ 
 
 $$ \ \ \ \ \ \ \ + \frac {g N} {2} \vert \psi \vert ^4 + \frac {N} {2} \vert \psi \vert ^2 \int {V_g (\vert \vec{r} - \vec{u} \vert) \vert \psi (\vec{u},t)\vert ^2 d^3u}  \eqno(5)$$

Now we come to the gravitational potential energy $V_g$ appearing in eq.(5). In the framework of LED gravity, the hierarchy problem of the Standard Model can be ameliorated if (a) there exists a fundamental energy scale $M_*  c^2$ ($\approx$ 1-1000 TeV, orders of magnitude less than the Planck energy $\cong  10^{19}$ GeV) for all interactions, and (b) there are additional sub-mm scale spatial dimensions, so that the perceived weakness of Newtonian gravity on large scales in (3+1)-dimensional space-time is due to the gravitational field lines spilling into the hidden spatial dimensions \cite{bec1,bec2}. In this formalism, the gravitational potential energy $V_g (r) $ between two point masses $m_1 $ and $m_2$ separated by a distance $r$ is given by,
$$ V_g (r) \cong  - \frac {m_1 m_2} {m^2_{pl}}\frac{\hbar c} {r}, \ \ \ \ \ r \gg R_* $$
$$\ \ \ \ \ \ \approx  - \frac {(R_*(n))^n m_1 m_2} {m^2_{pl}} \frac{\hbar c} {r^{n+1}}, \ \ \ \ \ r \ll R_* \eqno(6)$$
where $m_{pl} \equiv \sqrt{\frac {\hbar c} {G}} $ is the Planck mass corresponding to the standard Newton's gravitational constant G  and $R_* (n)$ is the radius of the extra dimensional n-torus given by,
$$R_* (n)=\bigg (\frac {m_{pl}} {M_*} \bigg )^{2/n} \frac {\hbar } {2 \pi M_* c} \eqno(7)$$
for $n$ =1,2,... According to eq.(6), closer one probes stronger is the gravity  on scales smaller than $R_* (n)$. In the next section, we examine its implications on low lying excitations of BEC.

\section{Variational method, gravity and BEC oscillation modes}

 Solving eq.(3) with $V_g$ given by eq.(6) is a nontrivial task. Instead, we take recourse to a variational method developed to study stability and low energy excitations of BEC \cite{bec8,bec9,bec10}. In this approach, the parameters of a trial wavefunction $\psi_{tr}$ is obtained by demanding that the action  is extremized by $\psi_{tr}$.  Since attractive contact  interactions (i.e. $a < 0$) is known to cause collapse of BEC \cite{bec9,bec11} for sufficiently large N, stability analysis with  gravitational interactions included is worth studying. 
 
 For this purpose, we consider  a spherically symmetric trap potential,
 $$V_{ext}= \frac{1}{2} m w^2_0 \ r^2\eqno(8)$$
  and choose  a  normalized trial wavefunction \cite{bec9},
$$\psi_{tr} (\vec{r},t)= A(t)\exp{(-r^2/2\sigma^2(t))} \exp {(i B(t) r^2)}\eqno(9)$$
where $A(t)$, $\sigma (t)$ and $B(t)$ are amplitude, width and phase parameters, respectively, that need to be determined from extremization of the action (eqs.(4) and (5)). As $\psi_{tr}$ is normalized, $A(t)$ and $\sigma (t)$ are related by,
$$\vert A(t) \vert^2= (\sqrt{\pi} \sigma(t))^{-3}\eqno(10)$$
so that,
$$A(t)= (\sqrt{\pi} \sigma(t))^{-3/2} \exp{(i \gamma (t))}\eqno(11)$$
where $\gamma(t)$ is a time dependent phase.
Substitution of eqs.(8)-(11) in eq.(5) and  carrying out the spatial integral thereafter lead to the  following Lagrangian,
$$L= \int {d^3r \ \mathcal{L}} =  \hbar \dot{\gamma}  + L_{int} + \frac {g N} {4\sqrt{2} \pi^{3/2} \sigma^3} + $$

$$\ \ \ \ \ \ \ \ + \frac{3}{2} \sigma^2 \bigg [\hbar \dot{B} + \frac{2 \hbar^2}{m} B^2 + \frac{\hbar^2}{2 m \sigma^4} + \frac{1}{2} m w^2_0 \bigg ] \eqno(12)$$
where the gravity term is,
$$L_{int} \equiv \frac {N} {2} \int {d^3r \vert \psi (\vec{r},t) \vert ^2 \int {V_g (\vert \vec{r} - \vec{u} \vert) \vert \psi (\vec{u},t)\vert ^2 d^3u}}  \eqno(13) $$

Using eq.(6) for $V_g$ the above integral can be evaluated analytically for $n=0$ and $n=1$ cases  so that,
$$L_{int} = - \frac {\alpha_0} {\sigma} \ \ \ \ \ \ \ \ \mbox{for} \ \ n=0 \eqno(14a)$$
$$\ \ \ \ \ \ = - \frac {\alpha_1} {\sigma^2} \ \ \ \ \ \ \ \ \mbox{for}\ \  n=1 \eqno(14b)$$
where,
$$\alpha_0 \equiv \frac {N \hbar c} {\sqrt{2 \pi}} \bigg (\frac {m} {m_{pl}} \bigg )^2 \ \ \ ,\eqno(15a)$$
$$\alpha_1 \equiv \frac {N R_* \hbar c} {2\sqrt{2}} \bigg (\frac {m} {m_{pl}} \bigg )^2 \sum^\infty_{k=0} \frac {1}{k!} \frac {(2k+1)!!}{2^{2k}(2k+1)}$$
$$\ \ \ \ \ \ \cong \frac {N R_* \hbar c} {2} \bigg (\frac {m} {m_{pl}} \bigg )^2\eqno(15b) $$
Extremizating the action entails  Euler-Lagrange equations  $\frac{d}{dt} (\partial L/\partial \dot{q_j}) - (\partial L/\partial q_j)=0$, for j=1 and 2, with $q_1 \equiv B$, $q_2\equiv \sigma $ and $L$ given by eq.(12) ($\gamma (t) $ is non-dynamical as it  appears only as an additive total derivative term in eq.(12)). The equations of motion are,
$$B(t)=\frac{m}{2\hbar} \frac{\dot{\sigma}}{\sigma} \eqno(16a)$$
and,
$$\hbar \dot{B} + \frac{2 \hbar^2}{m} B^2 - \frac{\hbar^2}{2 m \sigma^4} - \frac {g N} {4\sqrt{2} \pi^{3/2} \sigma^5} + \frac{1}{2} m w^2_0 = - f_n(\sigma)\eqno(16b)$$
where,
$$f_n(\sigma)= \frac{\alpha_0}{3 \sigma^3}\ \ \ \ \ \mbox{for}\ n=0$$
$$\ \ \ \ \ \ \ \ = \frac{2 \alpha_1}{3 \sigma^4}\ \ \ \ \ \mbox{for}\ n=1 \eqno(16c)$$
By combining eqs.(16a) and (16b), one arrives at the relevant equation needed to study small amplitude oscillations in an ultra-cold cloud of bosons,
$$m \ddot{\sigma}= - m w^2_0 \sigma  + \frac{\hbar^2}{ m \sigma^3} +  \frac {g N} {2\sqrt{2}\pi^{3/2} \sigma^4} - 2 \sigma f_n(\sigma) \eqno(17a)$$
Employing the following dimensionless quantities \cite{bec9} that make use of the BEC ground state scale $\sqrt{\hbar/m w_0}$,
$$ v \equiv \frac{\sigma} {\sqrt{\hbar/m w_0}}\ ,\ \ \tau \equiv w_0 t, \ \ \ P \equiv \sqrt{\frac{2}{\pi}}\frac{Na}{\sqrt{\hbar/m w_0}} $$ $$\Rightarrow \frac {g N} {2\sqrt{2}\pi^{3/2} m}=\frac{\hbar^2}{m^2} \sqrt{\hbar/m w_0}\  P \ \ \ ,\eqno(17b)$$
along with eq.(16c) in eq.(17a), we obtain,
$$\frac{d^2 v}{d\tau^2} = -v + \frac{1}{v^3} + \frac{P}{v^4} + F_n(v)\eqno(18)$$
for $n=0,1$, where the dimensionless gravitational accelerations have the forms,
$$F_0(v)= - \sqrt{\frac{2}{3 \pi^2}} \ N \bigg (\frac{m}{m_{pl}} \bigg )^2 \bigg (\frac{c}{w_0 \sqrt{\hbar/m w_0} } \bigg )\  v^{-2} \eqno(19a)$$
$$\ \mbox{and},\ \ F_1(v)= - \frac{2}{3} N \bigg (\frac{m}{m_{pl}} \bigg )^2 \bigg (\frac{R_*}{\hbar/m c} \bigg )\ v^{-3} \eqno(19b)$$
The RHS of eq.(18) corresponds to an effective potential $\Phi_n(v) \  (n=0,1) $ given by,
$$\Phi_n(v)= \frac{1}{2} \bigg [v^2  + \frac{1}{v^2} \bigg ] + \frac{P}{3  v^3} + v F_0(v)\ \                 \mbox{for} \ \ n=0\eqno(19c)$$
$$\ \ \ \ \ \ \ \ \ \ \ \ \ =\frac{1}{2} \bigg [v^2  + \frac{1}{v^2} \bigg ] + \frac{P}{3  v^3} + \frac{v F_1(v)}{2}\ \ \mbox{for}\ \ \ n=1\eqno(19d)$$  
In order to study small amplitude oscillation modes, one needs to find the minima of $\Phi_n(v)$. So, from $ \Phi^\prime _n(v)=0 $, the task here boils   down to determining the zeroes of the  quintic polynomial,
$$ v^5 \ - \ v \ - \ P \ - \ v^4 \ F_n(v)=0 \eqno(20)$$         
To estimate numerically the real positive roots $v_0$ of eq.(20) and the excitation frequencies proportional to $\sqrt{\Phi^{\prime \prime}_n(v_0)}$, we make use of  typical experimental length scales, 
$$\sqrt{\hbar/m w_0} \approx 10^{-4}\ \mbox{cm};\ (c/w_0)(m/m_{pl})^2 \approx 3\times 10^{-26} \ \mbox{cm};$$
$$a\approx 10^{-6} \mbox{cm} ;\  \ R_*(1) \approx 200\ \mu\mbox{m} ; \ \ \hbar/(m c)\approx 1.6 \times 10^{-16} \ \mbox{cm}\ , \eqno(21)$$
having in mind a BEC comprising of ${ }^{133}\mbox{Cs}$ for which $(m/m_{pl})^2 \cong 4.9\times 10^{-34}$.

Since both $P$ and $ -F_n(v) $ increase with $N$ with the latter being negligibly smaller by orders of magnitude due to the smallness of $(m/m_{pl})^2 $ inspite of the other factors (see eqs.(17b),(19a,b)  and (21)), it is obvious that the s-wave scatterings completely swamp the gravitational corrections to the excitation frequencies. The oscillation modes of such a problem in the absence of gravity had already been studied by  Perez-Garcia et al. \cite{bec9}

To circumvent the dominance of binary s-wave scattering one may, along with augmenting N,  invoke  Feshbach resonance \cite{bec13,bec14,bec15,bec16}. This effect enables experimentalists to  tune the scattering length $a$  magnetically, and reduce it to zero. Hence, with a  vanishing $ P $, in the $n=0$ case (i.e. pure Newtonian gravity), one finds that for $N < 10^{21}$, the real positive root $v_0$  of eq.(20) is very close to unity corresponding to a frequency of $\omega = 2 w_0$, as though the presence of $ F_0(v) $ did not matter.

However, for  macroscopically large occupation numbers $N= 10^{22}\ \mbox{and}\ 10^{23}$ (BECs of future), one finds significant departures:  $v_0=0.78 $, $\omega=2.4 w_0$  and  $v_0=0.12 $, $\omega=66 w_0 $, respectively. Because of the $3/v^4_0$ term in $\Phi^{\prime \prime}_n(v_0)$, one expects a higher excitation frequency as $v_0$ becomes smaller than unity.
Although these results suggest that rise in self-gravity due to increase in N beyond $10^{22}$ makes the ultra-cold gas cloud  shrink  drastically, caution needs to be exercised in concluding so. For, when the number density $\approx N (\sqrt{\hbar/m w_0}\  v_0)^{-3}$ becomes very large, other subatomic effects will start dominating  and, also, it is likely that the variational method demands more care in such circumstances. For instance, when $N= 10^{22}$, our result $v_0=0.78 $ implies a mean separation between atoms in the condensate to be  about $10^{-11} $ cm! Nevertheless, the observed pathology for $N\geq 10^{22}$ situation suggests that it would be interesting  to study the numerical solutions of GPE, 
 with Newtonian gravity added, for macroscopic BEC.
 
 There is  another way of getting around the problem of high density for large occupation numbers. One could  increase the length scale $\sqrt{\hbar/(m w_0)}$ by choosing a weaker trap potential. Hence, to ensure  mean separation not to fall below 10 Angstroms, the trap frequency $w_0$ must satisfy the condition,
 $$ w_0 < 10^{14} v^2_0 \bigg (\frac{3 N}{4\pi} \bigg )^{-2/3} \bigg (\frac{\ \ \hbar/m\ \ } {\mbox{cgs units}} \bigg )\ ,$$
 where $v_0$ is the positive root of eq.(20) corresponding to the Newtonian gravity case.
 
In the n=1 case ($1/r^3$ gravity), when $P=0$, the  non-zero roots of  eq.(20) satisfy,
$$v^4_0= 1 - (2N/3)(m/m_{pl})^2 (R_*/(\hbar/mc))\eqno(22)$$
implying that the roots are complex when,
$$ N > N_{cr}\equiv(3/2)(m/m_{pl})^{-2} (R_*/(\hbar/mc))^{-1}\eqno(23)$$
This is easily understood given that the potential $\Phi_1 $ of eq.(19d) can be expressed as,
$$\Phi_1(v)= \frac{1}{2}v^2 +   \frac{1}{2v^2} \bigg [ 1 - \frac{N}{N_{cr}} \bigg ] \eqno(24) $$
 provided $a$ has been magnetically tuned to zero. From eq.(24), it is clear that the potential is no longer bounded from below when the occupation number exceeds  $N_{cr}$.
 
  This signals instability for the BEC since its size characterized by $\sigma (t)$ rolls down towards 0 as it tries to lower its potential energy. From the  values provided in eq.(21), the onset of instability starts at $N_{cr}= 2.4 \times 10^{19}$. While, if  $R_*(1)$ is smaller $ \approx 1 \mu \mbox{m}$, the critical occupation number for ${ }^{133}\mbox{Cs}$ rises to $\approx 5 \times 10^{21}$.
 However, when $N<N_{cr}$, there is one positive root of eq.(22), and the corresponding excitation frequency is  $2w_0$, albeit independent of $n=1$ gravity.
\section{Conclusion} 
 Within the ambit of variational method, we have found  that occupation numbers in excess of $N_{cr}$     cause collapse of BEC for attractive gravity falling off as $r^{-3}$. This can be subjected to experimental verification only when one attains BECs with macroscopically large occupational numbers $\approx 10^{19}$ - $ 10^{22}$. For  higher values of $N$, even Newtonian gravity appears to have significant effect on the BEC dynamics that needs to be studied more carefully. The consequences of $n\geq 2$ LED  theories on BEC, though not  covered in this paper, need to be studied. In particular, it would be interesting to see whether their effects could be disentangled from those arising from other atomic interactions  like van der Waals force. 
\begin{acknowledgments}
It is a pleasure to thank N. D. Hari Dass, Michel Devoret, Romesh Kaul and T. R. Govindarajan for stimulating discussions.
\end{acknowledgments}


\end{document}